\documentclass[reprint,superscriptaddress,showpacs,preprintnumbers,amsmath,amssymb,aps,prb]{revtex4-1}
\usepackage{graphicx}
\usepackage{color}
\usepackage{braket}

\begin{document}

\title{Inversion of the exciton built-in dipole moment in In(Ga)As quantum dots via nonlinear piezoelectric effect}

\author{Johannes Aberl}
\email[]{johannes.aberl@jku.at}
\affiliation{Institute of Semiconductor and Solid State Physics, Johannes Kepler University Linz, Altenbergerstra{\ss}e 69, A-4040 Linz, Austria}

\author{Petr Klenovsk\'y}
\email[]{klenovsky@physics.muni.cz}
\affiliation{Department of Condensed Matter Physics, Masaryk University, Kotl\'a\v{r}sk\'a, CZ-61137 Brno,
Czech Republic}
\affiliation{Central European Institute of Technology, Masaryk University, Kamenice 753/5, CZ-62500~Brno, Czech~Republic}

\author{Johannes S. Wildmann}
\affiliation{Institute of Semiconductor and Solid State Physics, Johannes Kepler University Linz, Altenbergerstra{\ss}e 69, A-4040 Linz, Austria}

\author{Javier Mart\'in-S\'anchez}
\affiliation{Institute of Semiconductor and Solid State Physics, Johannes Kepler University Linz, Altenbergerstra{\ss}e 69, A-4040 Linz, Austria}

\author{Thomas Fromherz}
\affiliation{Institute of Semiconductor and Solid State Physics, Johannes Kepler University Linz, Altenbergerstra{\ss}e 69, A-4040 Linz, Austria}

\author{Eugenio Zallo}
\affiliation{Institute for Integrative Nanosciences, IFW Dresden, Helmholtzstra{\ss}e 20, D-01069 Dresden, Germany}
\affiliation{Paul-Drude-Institut f{\"u}r Festk{\"o}rperelektronik, Hausvogteilplatz 5-7, 10117 Berlin, Germany}

\author{Josef Huml\'i\v{c}ek}
\affiliation{Department of Condensed Matter Physics, Masaryk University, Kotl\'a\v{r}sk\'a, CZ-61137 Brno,
Czech Republic}
\affiliation{Central European Institute of Technology, Masaryk University, Kamenice 753/5, CZ-62500~Brno, Czech~Republic}

\author{Armando Rastelli}
\affiliation{Institute of Semiconductor and Solid State Physics, Johannes Kepler University Linz, Altenbergerstra{\ss}e 69, A-4040 Linz, Austria}

\author{Rinaldo Trotta}
\email[]{rinaldo.trotta@jku.at}
\affiliation{Institute of Semiconductor and Solid State Physics, Johannes Kepler University Linz, Altenbergerstra{\ss}e 69, A-4040 Linz, Austria}

\date{\today}

\begin{abstract}
We show that anisotropic biaxial stress can be used to tune the built-in dipole moment of excitons confined in In(Ga)As quantum dots up to complete erasure of its magnitude and inversion of its sign. We demonstrate that this phenomenon is due to piezoelectricity. We present a model to calculate the applied stress, taking advantage of the so-called piezotronic effect, which produces significant changes in the current-voltage characteristics of the strained diode-membranes containing the quantum dots. Finally, self-consistent $\mathbf{k}\cdot\mathbf{p}$ calculations reveal that the experimental findings can be only accounted for by the nonlinear piezoelectric effect, whose importance in quantum dot physics has been theoretically recognized although it has proven difficult to single out experimentally.
\end{abstract}


\maketitle

\section{Introduction}

Semiconductor quantum dots (QDs) are currently emerging as one of the most promising sources of nonclassical light on which to base future quantum technologies \cite{Aharonovich:16}. This success is in large part due to the outstanding experimental and theoretical work on QD physics that has been carried out over the last decades. These studies have not only enabled a detailed understanding of the fundamental properties of these ``artificial atoms", but have also offered the means to tailor their interactions with the environment, which is the key to make them suitable for envisioned applications.\\
In spite of these accomplishments, however, the extreme sensitivity of the electronic properties of QDs to tiny variations of their shape, size, composition, built-in strain fields, as well as to external perturbations \cite{Trotta:13} very often makes it difficult to single out the physical effects which are responsible for particular experimental observations. This is especially true for statistical studies performed on dissimilar QDs that aim at grasping general trends applicable to all of them. To explain this point further, we focus on the effect of piezoelectric fields, whose importance in theoretical semiconductor physics is well documented \cite{Bester:06,Migliorato:06,Migliorato:14,Caro:15}. Seminal works have demonstrated that in conventional III-V QDs grown on (100) substrates, first- and second-order contributions to the piezoelectric field tend to oppose each other so that its total effect on the QD properties is found to be small \cite{Bester:06_2,Schliwa:09}. This is clearly not the case for GaN QDs \cite{Winkelnkemper:06,Honig:13} and CdSe nanocrystals \cite{Segarra:16}, but for In(Ga)As QDs, which are usually grown on nonpolar GaAs(001) substrates and are of interest for quantum optics, piezoelectricity is very often neglected \cite{Gong:08,Zielinski:13}.\\
From the experimental side it is not straightforward to recognize its importance. Despite the weak effect on the energy of the states, piezoelectricity is expected to have a strong influence on the position and shape of the electron and hole wave functions \cite{Bester:06_2} and, in turn, on the sign and magnitude of the exciton ($X$) built-in dipole moment. Early experiments \cite{Fry:00} showed that the sign of the built-in dipole moment is inverted with respect to the predicted sign \cite{Grundmann:95}, a fact which was mainly explained by the shape and composition of the investigated QDs. In contrast, subsequent studies performed with In(Ga)As QDs grown on high-index substrates with polar orientation \cite{Levin:01} demonstrated that the observed electron-hole alignment is not a general feature, and that piezoelectricity has to be taken into account for a correct interpretation of the experimental results, as confirmed very recently \cite{Germanis:16}. Related experiments have demonstrated that the sign of the $X$ dipole can even vary for different QDs within the same sample \cite{Bennett:10}. It is therefore rather evident that it is experimentally challenging to single out the role of piezoelectricity in QDs.\\
\begin{figure}[h]
\includegraphics[scale=1]{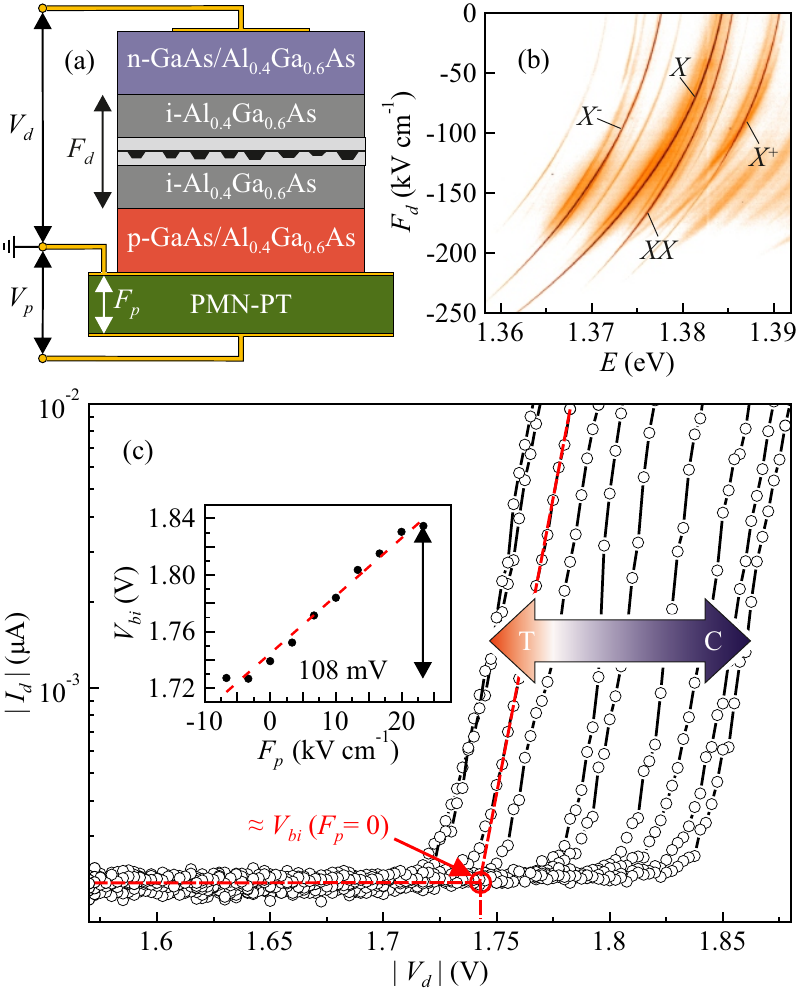}
\caption{\label{Fig 1} (a) Sketch of the p-i-n diode bonded onto a PMN-PT piezoelectric substrate. The investigated In(Ga)As QDs were embedded in the central i-GaAs layer (light grey). The emission properties of the QDs can be tuned by using voltages applied to PMN-PT ($V_p$) and diode ($V_d$). The color-coded $\mu$-PL map in panel (b) shows the transition energies of exciton ($X$), biexciton ($XX$), negative and positive trions ($X^-$, $X^+$) of a single QD as a function of the electric field across the diode $F_d$ (for $F_p=0$). (c) Shift of the $I\text{-}V$ traces ($I_d$ and $V_d$ are negative in the displayed range) of the diode in response to applied stress (tensile, T or compressive, C). The built-in voltage $V_{bi}$ was estimated from the intersection of the fitted part of the forward bias region with the saturation current (red line). The shift of $V_{bi}$ for different steps of $F_p$ is ascribed to the piezotronic effect and is provided in the inset.}
\end{figure}
In this paper, we demonstrate that externally induced anisotropic strain fields can be used for wave function engineering and to even force an inversion of the exciton built-in dipole moment in the very same In(Ga)As QD. We achieve this result by integrating light-emitting-diode (LED) nanomembranes onto a piezoelectric actuator [see Fig.~\ref{Fig 1}(a)] capable of delivering variable strain fields \cite{Trotta:12,Trotta:12_2}. Differently from previous results \cite{Kuklewicz:12} revealing a strain-dependence of the $X$ dipole moment, we demonstrate that its tuning (and inversion) is driven by the piezoelectric effect. The induced piezoelectric field was estimated by exploiting a piezotronic-like phenomenon in the LED used \cite{Zhang:11}, i.e., a sizable strain-induced shift of its current-voltage ($I\text{-}V$) characteristics. Finally, self-consistent $\mathbf{k}\cdot\mathbf{p}$ calculations reveal that the inversion of the dipole moment is dominated by the nonlinear terms of the piezoelectric field.
\section{Experimental Methods}
The microphotoluminescence ($\mu\text{-PL}$) measurements were performed at low temperatures ($8\,\mathrm{K}$) by using a helium flow cryostat. A femtosecond Ti-sapphire-laser (operated at $850\,\mathrm{nm}$) was focused by a microscope objective ($0.42$ numerical aperture) to address single QDs. The $\mu$-PL spectra were recorded via a spectrometer connected to a liquid-nitrogen-cooled charge-coupled device. Polarization-resolved measurements were performed using a combination of a rotatable $\mathrm{\lambda/2}$ wave plate and a fixed linear polarizer placed in front of the spectrometer in order to identify the origin of the transitions in the $\mu\text{-PL}$ spectra [see Fig.~\ref{Fig 1}(b)] and to estimate the exciton fine-structure splitting (FSS). The p-i-n diode nanomembrane containing In(Ga)As QDs was grown by molecular beam epitaxy. The nanomembrane was transferred via a flip-chip process and bonded onto a $[\mathrm{Pb}[\mathrm{Mg}_{1/3}\mathrm{Nb}_{2/3}]\mathrm{O}_{3}]_{0.72}\text{-}[\mathrm{PbTiO}_{3}]_{0.28}$ (PMN-PT) piezoelectric actuator by gold thermo-compression bonding (for further details on the sample structure and device fabrication see Refs.~\onlinecite{Trotta:12,Trotta:12_2}). The emission properties of the QDs can be varied using two ``tuning knobs": stress and electric fields, which are applied via voltages $V_p$ and $V_d$, respectively. A positive (negative) voltage $V_p$ applied to the PMN-PT induces a compressive (tensile) anisotropic biaxial stress in the nanomembrane. The corresponding electric field across the PMN-PT is given by $F_p=V_p/d_p$, where $d_p\approx 300\,\mathrm{\mu m}$ is its thickness. The electric field across the LED is instead given by $F_d=-(V_d+V_{bi})/d_d$, where $-1.9\,\mathrm{V}<V_d<1.9\,\mathrm{V}$ is the applied voltage, $V_{bi}$ the diode's built-in voltage (positive) and $d_d\approx 150\,\mathrm{nm}$ the thickness of the intrinsic region. For operation in reverse bias, $F_d$ is oriented parallel to the QD's growth direction, i.e. from top to bottom in Fig.~\ref{Fig 1}(a). Increasing the magnitude of $F_d$ leads to a redshift of all transitions, as shown in Fig.~\ref{Fig 1}(b).
\section{Tuning of the $X$ electric dipole moment}
\begin{figure}[h]
\includegraphics[scale=1]{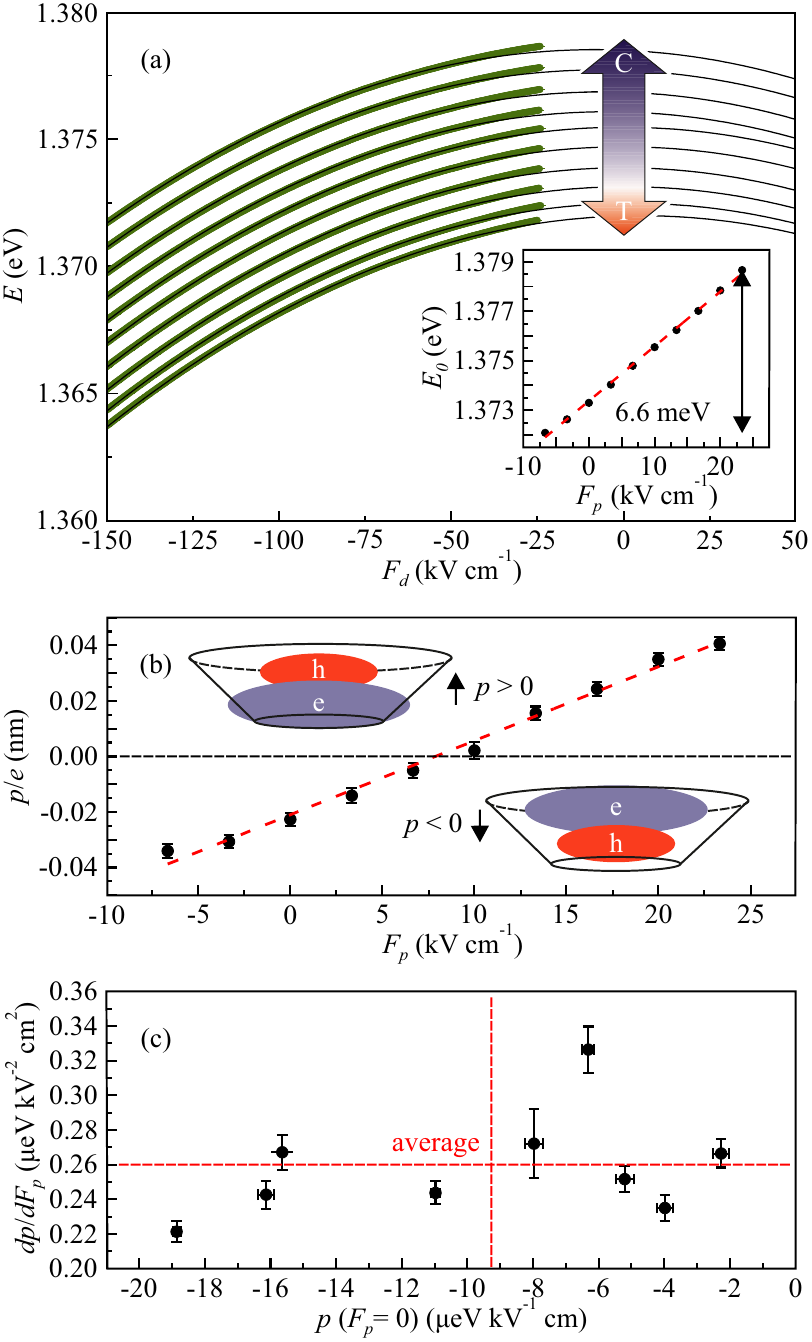}
\caption{\label{Fig 2} (a) Determination of the stress dependence of $E_0$, $p$ and $\beta$ for voltages $V_p$ in the range from $V_p=-200\,\mathrm{V}$ up to $V_p=700\,\mathrm{V}$. The measured data (green points) are fitted with Eq.~(\ref{Eq 1}) (black lines) to extract $E_0$, $p$ and $\beta$. In the inset, the linear shift of $E_0$ with $F_p$ is shown. The change in carrier separation $p/e$ of $X$ is shown in panel (b). The sketches illustrate the strain-induced inversion of the alignment of the electron and hole wave functions. The values obtained for $p$ at $F_p=0$ for all measured excitons are provided in panel (c) along with the corresponding tuning rates $dp/dF_p$.}
\end{figure}
We now show how strain can be used to invert the $X$ built-in dipole moment. We start out by fitting the measured Stark shift of the $X$ transition via
\begin{equation}
E=E_0-p F_d+\beta F_d^2
\label{Eq 1}
\end{equation}
where $E_0$ is the transition energy at $F_d=0$, $p=ez$ the built-in dipole moment ($e$ elementary charge, $z$ is the carrier separation along the growth direction), and $\beta$ the polarizability \cite{Finley:04}. Repeating this procedure for each value of $F_p$ allows us to extract the stress dependence of $E_0$, $p$ and $\beta$, as illustrated in Fig.~\ref{Fig 2}(a). Obviously, the determination of these parameters is tightly connected to the use of correct values of $F_d$ and therefore of $V_{bi}$, which can be estimated from the $I\text{-}V$ trace of the diode [see Fig.~\ref{Fig 1}(c)]. Herein it turned out that there is a substantial shift of $V_{bi}$ with applied stress of up to $\Delta V_{bi}= 108\,\mathrm{mV}$ (for $\Delta V_{p}=900\,\mathrm{V}$) which is about an order of magnitude larger than the expected shift produced by the strain-induced changes of the energy band gap ($\sim10\,\mathrm{meV}$). This effect, which we attribute to piezoelectricity, is of great importance not only for the data evaluation, but also for the theoretical model discussed below.\\
In Fig.~\ref{Fig 2}(b) we report the strain-dependence of $p/e$ for the $X$ confined in one of the nine measured QDs. The dipole moment shifts almost linearly with $F_p$, i.e. with applied stress, with an average tuning range of $\braket{\Delta p}/e=(0.071\pm 0.007)\,\mathrm{nm}$ and an average slope of $\braket{dp/dF_p}=(0.26\pm 0.04)\,\mathrm{\mu eV}\,\mathrm{kV^{-2}}\,\mathrm{cm^2}$. Most importantly, the applied stress is sufficient to suppress the electric dipole and invert its sign, i.e. \textit{to swap the position of electron and hole wave functions} inside the QD. The inversion of $p$ has been observed in four out of the nine measured QDs, and it is mainly determined by the value of $p$ for $F_p=0$, see Fig.~\ref{Fig 2}(c). This is found to be always negative, i.e. at $F_p=0$ the hole (electron) tends to be located closer to the QD apex (base). The measured (linear) strain-induced shift of the zero-field transition energy $E_0$ [$\braket{dE_0/dF_p}=(0.21\pm 0.02)\,\mathrm{\mu eV}\,\mathrm{kV^{-1}}\,\mathrm{cm}$] is in good agreement with previous works~\cite{Trotta:12,Trotta:13}. The full set of data including $E_0$, $p$ and $\beta$ for all transitions can be found in the Supplemental Material \cite{SupMat}.
\section{Determination of stress configuration}
\begin{figure}
\includegraphics[scale=1]{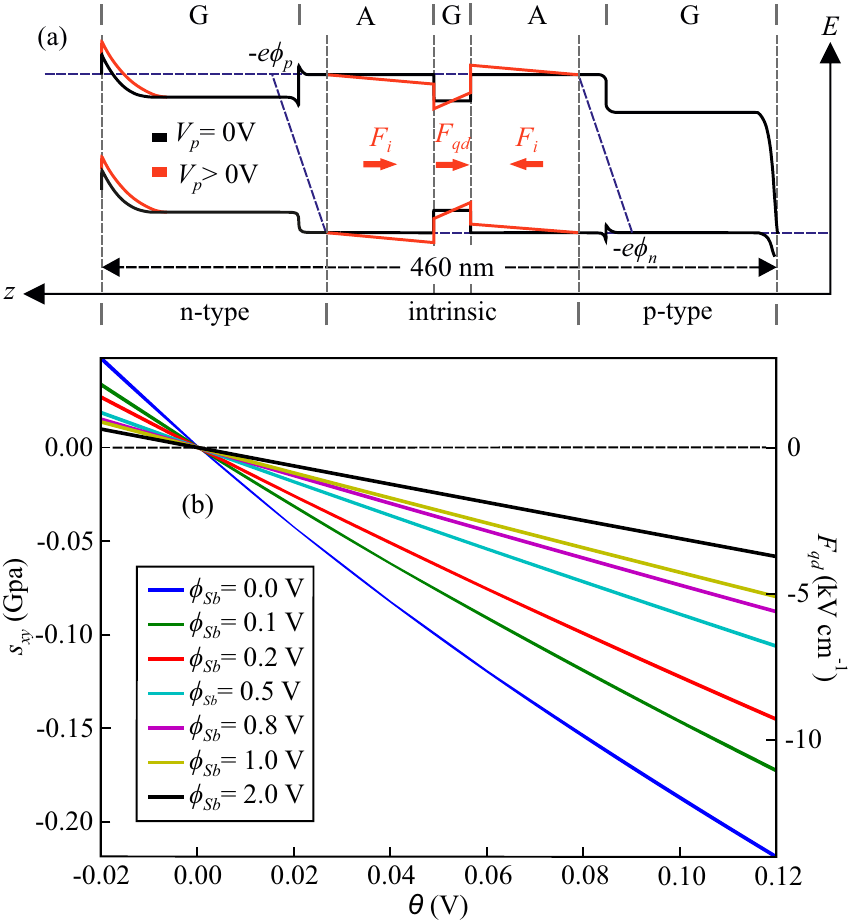}
\caption{\label{Fig 3} (a) Schematic band profile of the studied diode. Hereby $-e\phi_n$ and $-e\phi_p$ denote the quasi-Fermi-levels for electrons and holes, respectively. G and A simply label the different materials GaAs and $\mathrm{Al_{0.4}Ga_{0.6}As}$. The red curve indicates the strain-induced changes in the band alignment. In panel (b) the induced shear stress $s_{xy}$ in the GaAs layer containing the QDs as well as the resulting electric field $F_{\mathrm{qd}}$ are plotted as a function of $\theta$ for several (initial) values of the Schottky barrier $\phi_{Sb}$.}
\end{figure}

To explain the physics underlying the inversion of the $X$ built-in dipole moment, it is fundamental to gain knowledge of the type of in-plane stress delivered by the PMN-PT actuator. Any in-plane stress configuration can be described by the three independent components of the stress tensor ($s_{xx}$, $s_{yy}$ and $s_{xy}$) or, equivalently, by two principal stresses $S_1$, $S_2$ applied at an angle $\alpha$ with respect to a crystal axis ([100] in our case). The two sets of three independent parameters are related to each other via $s_{xx,yy}=\frac{S_1+S_2}{2}\pm\frac{S_1-S_2}{2}\,\mathrm{cos}(2\alpha)$ and $s_{xy}=\frac{S_1-S_2}{2}\,\mathrm{sin}(2\alpha)$. Therefore, an arbitrary in-plane stress can be fully characterized by $S_1-S_2$, $S_1+S_2$, and $\alpha$. This requires the knowledge of several observables as $F_p$ is varied. In our experiment we monitor (i) the shift of the $X$ transition energy $\Delta E_0$ [see inset of Fig.~\ref{Fig 2}(a)], (ii) the changes of the magnitude of the FSS along with the corresponding $X$ polarization angle, and (iii) the shift of the $I$-$V$ trace of the diode. Point (i) is related to the hydrostatic part of the stress given by $\Delta E_0=\widetilde{a}(S_1+S_2)$, where $\widetilde{a}$ is a parameter related to the elastic constants renormalized by the deformation potentials. Since $\widetilde{a}$ is known \cite{Trotta:15}, we can estimate $S_1+S_2$. Next, we use (ii) to estimate the direction $\alpha$ of the applied stress by using a recently developed model for the $X$ Hamiltonian \cite{Trotta:15,Trotta:16}. For the QDs investigated in this work we estimate $\alpha=55^{\circ}$ (note that $\alpha=45^{\circ}$ correspond to [110] direction), as discussed in the Supplemental Material \cite{SupMat}.
 Finally, we exploit (iii) to estimate $s_{xy}$ and, since $\alpha$ is known, $S_1-S_2$ (see later in the text). As the shift of the $I\text{-}V$ trace onset is a quite peculiar phenomenon, we discuss its origin below.\\
As mentioned, the measured shift of the $I\text{-}V$ trace cannot be accounted for by the strain-induced change of the i-$\mathrm{Al_{0.4}Ga_{0.6}As}$ band gap. In highly doped n-type GaAs the Fermi level ($E_F$) lies inside the conduction band \cite{Blakemore:82}, and there is a potential difference at the interface between a semiconductor and a metal stemming from their different work functions \cite{Ng:95}. Although there are several possibilities of the arrangement \cite{Ng:95,Cowley:65} we restrict ourselves (motivated by a typical scenario for n-type GaAs/Au interface \cite{Hudait:01}) to the case when a Schottky barrier $\phi_{Sb}$ is present at the Au contacts to the p-i-n diode \cite{Werner:91}. Under this condition, the observed $I\text{-}V$ shift stems from a strain-induced modification of $\phi_{Sb}$. More specifically, for $V_p\neq 0$ piezoelectric charges are generated at the edges of our structure as well as at every interface between two different materials. As illustrated via the band scheme in Fig.~\ref{Fig 3}(a) these charges produce an additional potential $\theta$ at the Au contacts, which effectively changes the current onset in the $I$-$V$ trace \cite{Zhang:11}. Furthermore, an additional (net) electric field $F_{qd}$ is created in the inner i-GaAs layer which acts on the QDs hosted therein. The magnitude and direction of $F_{qd}$ depends on the applied stress (T or C) as well as on the arrangement of different materials \cite{Huang:16}. The surface (piezoelectric) charge density $\sigma_p$ at the n-GaAs/Au interface is linked to the shear stress via $\sigma_p=e^n_{14}\widetilde{S}_{44}s_{xy}$, where $e^n_{14}$ is the piezoelectric constant of n-GaAs and $\widetilde{S}_{44}$ is one of the elastic compliance constants. By obtaining a relation between $\sigma_p$ and $\theta$ we can therefore exploit the measured shift of the $I\text{-}V$ trace onset to calculate $s_{xy}$ (and therefore $S_1-S_2$). The obtained dependence of $s_{xy}$ on $\theta$ as well as of the additional electric field in the QD layer $F_{qd}$ that is related to piezoelectric charges is shown in Fig.~\ref{Fig 3}(b). The corresponding relations and their derivation are provided in the Supplemental Material \cite{SupMat}. The sensitivity of $s_{xy}$ and $F_{qd}$ on $\theta$ is evident, especially for the small values of $\phi_{Sb}$ typical for our structure at low temperature (we estimated $0.1\,\mathrm{V}$ for the doping concentration of $5\cdot 10^{18}\,\mathrm{cm^{-3}}$ present in the n-GaAs\cite{Hudait:01}).\\
The knowledge of the shear part of the stress allows us to calculate $S_1-S_2$ for each value of $F_p$ and, finally, magnitude, direction and anisotropy of the in-plane stress delivered by the PMN-PT. This is found to be highly anisotropic (with a ratio $|S_1|/|S_2|\approx 3.16$), applied at $55^{\circ}$ with respect to the [100] direction and with magnitudes ($S_1+S_2$) as high as $-180\,\mathrm{MPa}$ (see Supplemental Material \cite{SupMat}). While this stress anisotropy is not expected for the [001] piezo cut used in this work, it is a common feature reported in the literature \cite{Trotta:12,Kumar:14,Ziss:17}.\\
\section{Origin of the dipole inversion}
\begin{figure}
\includegraphics[scale=1]{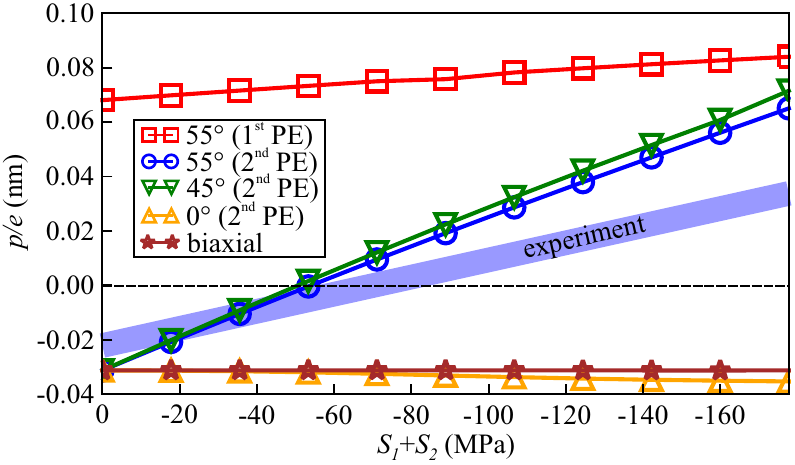}
\caption{\label{Fig 4} The dependence of $p/e$ on applied stress is shown for several configurations. The configuration estimated for our structure ($\alpha=55^{\circ}$) is plotted with (blue) and without (red) taking into account the second-order term of the piezoelectric field. Stress fields with same anisotropy and magnitude but with $S_1$ aligned along [110] (green, $\alpha=45^{\circ}$) and [100] (yellow, $\alpha=0^{\circ}$) directions and for the case of purely biaxial stress (brown) are also provided.}
\end{figure}

On this basis we performed calculations of the electronic structure of In(Ga)As QDs and investigate their dependence on the externally applied stress. The single-particle electronic levels were obtained by using the envelope function approximation based on the eight-band $\mathbf{k}\cdot\mathbf{p}$ method for electrons and holes employing the $\text{NEXTNANO}^3$ software package \cite{Birner:07}. The simulated QD was assumed to be of truncated cone shape with a radius of $20\,\mathrm{nm}$ ($10\,\mathrm{nm}$) at its base (apex) and a height of $3.5\,\mathrm{nm}$. The QD was embedded in a GaAs host and consisted of an $\mathrm{In_{x}Ga_{1-x}As}$ alloy with the In content linearly increasing from base ($\mathrm{x}=0.45$) to apex ($\mathrm{x}=0.8$) \cite{Wang:06}. The externally applied stress was simulated by changing the corresponding elements of the Bir-Pikus Hamiltonian \cite{Bir:74} and the calculations were performed in self-consistent Poisson-Schr\"{o}dinger equation loops.\\
The calculations furthermore account for the effect(s) induced by piezoelectric fields. Following the common definition, the piezoelectric response of a material is given in terms of the created polarization $\mathbf{P}$ which can be expanded as
\begin{equation}
P_i=\sum_{j=1}^{6}e_{ij}\epsilon_{j}+\frac{1}{2}\sum_{jk=1}^{6}B_{ijk}\epsilon_{j}\epsilon_{k}+\ldots
\label{Eq 2}
\end{equation}
where $\epsilon$ represents the independent components of the strain tensor in Voight notation (i.e. $\epsilon_1=\epsilon_{xx}$, $\epsilon_2=\epsilon_{yy}$, $\epsilon_3=\epsilon_{zz}$, $\epsilon_4=2\epsilon_{yz}$, $\epsilon_5=2\epsilon_{xz}$ and $\epsilon_6=2\epsilon_{xy}$) and $e$ and $B$ are the linear 
and nonlinear piezoelectric coefficients, respectively \cite{Wakata:11}. In zinc-blende crystals only four independent coefficients $e_{14}$, $B_{114}$, $B_{124}$ and $B_{156}$ are nonzero due to symmetry considerations \cite{Grimmer:07}. The importance of the different-order contributions to the piezoelectric polarization depends essentially on the magnitude of strain and its particular configuration in the considered material and/or structure. Differently from the diode model presented above, where the relatively small applied stresses (or equivalently strains) justify the use of the linear contribution only, the strain field around and inside the QD is at least an order of magnitude larger. Therefore the nonlinear response is expected to be strongly magnified and can even dominate \cite{Bester:06_2}. It is consequently reasonable to include at least the second-order contributions whereby the values for linear and quadratic piezoelectric coefficients were taken from Ref.~\onlinecite{Bester:06} (cubic terms are expected to be negligible \cite{Tse:13}).\\
In Fig.~\ref{Fig 4} we present the dependence of $p/e$ on $S_1+S_2$ for the stress configuration we estimated above with and without taking into account the second-order term of the piezoelectric field. Furthermore, we show the results obtained for stress fields with the same anisotropy and magnitude but with $S_1$ aligned along the [110] direction (i.e., the direction which maximizes the piezoelectric effect), the [100] direction (no piezoelectricity present), as well as for the case of purely biaxial stress (no piezoelectricity present).\\
A rich scenario can be observed. First, we notice that no appreciable variation of $p/e$ can be observed for $s_{xy}=0$, that is, in the absence of the piezoelectric effect (see the brown and the orange lines in Fig.~\ref{Fig 4}). In strong contrast, for the stress configuration estimated above (blue line) and for a similar one in which the piezoelectric field is maximized (green line), we observe a variation of the dipole moment comparable to the experimental observations. The discrepancy between the experimental and theoretical values of $p$ (and absolute value of $E_0$) is probably due to the specific QD shape considered for the theoretical calculations, which probably differs from the experimental one. It is also likely that the strain configuration we estimated is not exactly the one experienced by the specific QD due to strain inhomogeneities across the membranes.
Nonetheless, our calculations clearly show that \textit{the change in the dipole moment observed in the experiment can be only explained by the anisotropic strain configurations which switch on the piezoelectric effect}. Moreover, the large tuning rates achieved in the experiment can only be reproduced by including the second-order contribution to the piezoelectric field as apparent from the red line in Fig.~\ref{Fig 4}. Thus, the inversion of the exciton dipole moment we report in this work constitutes clear and rare experimental evidence of the importance of the nonlinear terms of the piezoelectric field in III-V QD systems.\\
It is worth emphasizing that, while in the calculations of Fig.~\ref{Fig 4} we used the values of the nonlinear piezoelectric coefficients reported in Ref.~\onlinecite{Bester:06} there are several works \cite{Migliorato:06,Wakata:11,Tse:13,Caro:15} that have reported different values for these coefficients. In the Supplemental Material \cite{SupMat} we discuss this point in detail and we show that the results of our calculations and our findings are preserved upon exchange of the values of the piezoelectric coefficients reported in Refs.~\onlinecite{Bester:06,Wakata:11,Tse:13,Caro:15}. We would also like to mention that we approximated the (experimental) absolute value of $p/e$ at $F_p=0$ in the performed calculations by theoretically considering a shear pre-stress of $s^{pre}_{xy}=200\,\mathrm{MPa}$. This pre-stress, which is already present in our device at $F_p=0$, is a common feature and is attributed to the bonding and poling process \cite{Kumar:14,Ziss:17}. However, it has no qualitative influence on the presented behavior of $p/e$ vs $S_1+S_2$, which, obviously, purely depends on the applied stress (which is estimated from the experimental data).\\
\section{Conclusion}
In conclusion, we have demonstrated experimentally and theoretically that piezoelectric fields can be used to engineer the wave function of excitons confined in In(Ga)As QDs and that, in this phenomenon, the nonlinear terms of the piezoelectric field dominate over the linear term. Our results are relevant not only for fundamental physics, because the effect of the piezoelectric field on the few-particle states in QDs can be now pinpointed from the experiments, but also for future applications. In fact, piezoelectricity can be used to modify the radiative lifetime, similar to vertical electric fields \cite{Bennett:10}. Moreover, the dipole moment can be engineered to limit the interaction of excitons with charges in the vicinity of the QD \cite{Callsen:15} or to modify the response of QDs to electric fields \cite{Salter:10}. In this context, it is worth noting that the tunability of the exciton dipole moment offered by the piezoelectric field is at least an order of magnitude larger than what can be obtained using magnetic fields \cite{Cao:15} and, in addition, can be achieved using a compact and scalable approach.
\begin{acknowledgments}
This work was financially supported by the European Research council (ERC) under the European Union's Horizon 2020 Research and Innovation Programme (SPQRel, Grant agreement No.~679183) and by the European Union Seventh Framework Programme 209 (FP7/2007-2013) under Grant Agreement No.~601126 210 (HANAS). P.~K. and J.~H. have been financially supported by the Ministry of Education, Youth and Sports of the Czech Republic under the project CEITEC 2020 (LQ1601). Furthermore, we acknowledge E.~Magerl, R.~Singh, G.~Bester, P.~Atkinson and O.~G.~Schmidt for fruitful discussions and support.
\end{acknowledgments}

\bibliography{dipole_inv_rev}

\end{document}